\documentclass[3p,times,twocolumn]{elsarticle}

\usepackage{ecrc}

\volume{00}

\firstpage{1}

\journalname{Nuclear Physics B Proceedings Supplement}

\runauth{B. Duclou\'e, T. Lappi, H. M\"antysaari}

\jid{nppp}

\jnltitlelogo{Nuclear Physics B Proceedings Supplement}

\usepackage{amssymb}

\usepackage[figuresright]{rotating}

\usepackage{amsmath}
\usepackage{hyperref}

\def\P{{\boldsymbol P}}
\def\k{{\boldsymbol k}}
\def\l{{\boldsymbol l}}
\def\p{{\boldsymbol p}}
\def\q{{\boldsymbol q}}

\newcommand{\der}{\mathrm{d}}
\newcommand{\xt}{{{\boldsymbol x}_\perp}}
\newcommand{\yt}{{{\boldsymbol y}_\perp}}
\newcommand{\bt}{{{\boldsymbol b}_\perp}}
\newcommand{\rt}{{{\boldsymbol r}_\perp}}
\newcommand{\kt}{{\k_\perp}}
\newcommand{\lt}{{\l_\perp}}
\newcommand{\pt}{{\p_\perp}}
\newcommand{\qt}{{\q_\perp}}

\newcommand{\ud}{\, \mathrm{d}}
\newcommand{\tr}{\, \mathrm{Tr} \, }
\newcommand{\nc}{{N_\mathrm{c}}}
\newcommand{\da}{d_\mathrm{A}}

\newcommand{\qso}{Q_\mathrm{s0}}

\newcommand{\lqcd}{\Lambda_{\mathrm{QCD}}}
\newcommand{\as}{\alpha_{\mathrm{s}}}

\newcommand{\Jpsi}{{J/\psi}}

\begin{document}

\begin{frontmatter}



\dochead{}

\title{Nuclear modification of forward $J/\psi$ production in proton-nucleus collisions at the LHC}


\author[jyu,hip]{B. Duclou\'e}
\author[jyu,hip]{T. Lappi}
\author[jyu]{H. M\"antysaari}
\address[jyu]{Department of Physics, University of Jyv\"askyl\"a, P.O. Box 35, 40014 University of Jyv\"askyl\"a, Finland}
\address[hip]{Helsinki Institute of Physics, P.O. Box 64, 00014 University of Helsinki, Finland}

\begin{abstract}
We re-evaluate the nuclear suppression of forward $J/\psi$ production at high energy in the Color Glass Condensate framework. We use the collinear approximation for the projectile proton probed at large $x$ and an up to date dipole cross section fitted to HERA data to describe the target in proton-proton collisions. We show that using the Glauber approach to generalize the proton dipole cross section to the case of a nucleus target leads to a nuclear modification factor much closer to LHC data than previous estimates using the same framework.
\end{abstract}


\end{frontmatter}


\section{Introduction}

The study of forward $\Jpsi$ production in high energy proton-proton and proton-nucleus collisions, which probes the target at very small $x$, can provide valuable information on gluon saturation. Indeed, the charm quark mass should be small enough to be sensitive to the saturation scale. On the other hand, it is large enough to provide a hard scale and thus to allow the use of a weak coupling treatment. It also has a clean experimental signature so this process has been the subject of many experimental studies to date.

In this work we will study $\Jpsi$ production at forward rapidities in proton-proton and proton-nucleus collisions at the LHC in the Color Glass Condensate (CGC) framework, using the color evaporation model (CEM) to treat hadronization. Since we work at forward rapidity, where the projectile is probed at large $x$ and the target at small $x$, we will use the ``hybrid model'' in which the projectile proton is treated as dilute and is described in terms of usual collinear parton distribution functions (PDFs). The collinear gluon emitted can then split into a $c\bar{c}$ pair either before or after the interaction with the target. These partons are then assumed to eikonally interact with the target, picking up a Wilson line factor in either the adjoint or the fundamental representation, depending on the particle.
The cross section for $c\bar{c}$ pair production is then described in terms of Wilson line correlators containing the information on the dense target. The same Wilson line correlators appear in calculations of other processes, such as total DIS cross sections, single and double inclusive particle production in proton-proton and proton-nucleus collisions, diffractive DIS and the initial state for hydrodynamical modeling of heavy ion collisions. This framework has thus a broad range of applications.

The modification of $\Jpsi$ production cross section in proton-nucleus compared to proton-proton collisions has been previously studied in the CGC framework~\cite{Fujii:2013gxa}. However it was found that the nuclear suppression predicted by this calculation was much stronger than measured later at the LHC.
In this work we re-evaluate this quantity in the same collinear ``hybrid'' framework, using a more careful treatment of nuclear geometry necessary to go from the description of a proton to the one of a nucleus in the CGC framework. This is motivated by the fact that it was observed for example in single inclusive light hadron production~\cite{Lappi:2013zma} that the disagreement of previous CGC calculations~\cite{Albacete:2010bs} with LHC data was mostly due to nuclear geometry effects. We also use the more recent dipole cross sections which were obtained in Ref.~\cite{Lappi:2013zma}.

\section{Formalism}
\label{sec:formulae}

In this work we use the simple color evaporation model (CEM) to describe the hadronization of the $c\bar{c}$ pair into a $\Jpsi$ meson. We note that it is also possible to treat hadronization in a more elaborate way, for example by using an expansion in terms of non-relativistic QCD as was done in Ref.~\cite{Ma:2015sia}, but here we focus on the importance of nuclear geometry. In the CEM a fixed fraction of the $c\bar{c}$ pairs, produced either in the color singlet or octet state, whose invariant mass is below the $D$-meson threshold, is assumed to hadronize into $\Jpsi$ mesons. The differential cross section with respect to the transverse momentum $\P_{\perp}$ and the rapidity $Y$ of the produced $\Jpsi$ reads
\begin{align} 
\frac{\ud\sigma_{\Jpsi}}{\ud^2\P_{\perp}\ud Y}
=
F_{\Jpsi} \; \int_{4m_c^2}^{4M_D^2} \ud M^2
\frac{\ud\sigma_{c\bar c}}
{\ud^2\P_{\perp} \ud Y \ud M^2}
\, ,
\label{eq:dsigmajpsi}
\end{align}
Where $m_c$ is the charm quark mass, $m_D=1.864$ GeV is the $D$ meson mass and $\frac{\ud\sigma_{c\bar c}}{\ud^2\P_{\perp} \ud Y \ud M^2}$ is the cross section for $c\bar{c}$ pair production with transverse momentum $\P_{\perp}$, rapidity $Y$ and invariant mass $M$. The nonperturbative constant $F_{\Jpsi}$ in Eq.~\ref{eq:dsigmajpsi} is related to the probability for a $c\bar{c}$ pair to transition to a $\Jpsi$.
In this work we will be mostly interested in the nuclear modification factor $R_\text{pA}$, defined as
\begin{align}\label{eq:defrpa}
R_\text{pA}= \frac{1}{A}\frac{\left . \ud \sigma/ \ud^2 
	\P_\perp \ud Y \right |_\text{pA}}
{\left . \ud\sigma/\ud^2 \P_\perp \ud Y \right |_\text{pp}} \; ,
\end{align}
where $\left . \ud\sigma/\ud^2 \P_\perp \ud Y \right |_\text{pp}$ and $\left . \ud \sigma/ \ud^2 
\P_\perp \ud Y \right |_\text{pA}$ are the cross sections in proton-proton and proton-nucleus collisions respectively. Therefore for this observable $F_{\Jpsi}$ plays no role and we don't need to fix it.

The formalism for gluon and quark pair production in the dilute-dense limit of the CGC has been studied in detail in Refs.~\cite{Blaizot:2004wu,Blaizot:2004wv} (see also Ref.~\cite{Kharzeev:2012py}) and used in several works, such as~\cite{Fujii:2005rm,Fujii:2006ab,Fujii:2013gxa,Fujii:2013yja}. In this framework, when using the collinear approximation to describe the gluon emitted by the projectile, the cross section for $c\bar{c}$ pair production reads, in the large-$\nc$ limit~\cite{Fujii:2013gxa}:
\begin{multline}
	\!\!\!\!\! \frac{\ud \sigma_{c\bar{c}}}{\ud^2\pt \ud^2\qt \ud y_p \ud y_q}
	= 
	\frac{\as^2 \nc}{8\pi^2 \da}
	\frac{1}{(2\pi)^2}
	\!\!\int\limits_{\k_\perp}\!
	\frac{\Xi_{\rm coll}(\pt + \qt,\k_{\perp})}{(\pt + \qt)^2}
	\\ \times
	\phi_{y_2=\ln{\frac{1}{x_2}}}^{q\bar{q},g}(\pt + \qt,\k_\perp)
	\;
	x_1 G_p(x_1,Q^2) \; ,
	\label{eq:dsigmaccbarcoll}
\end{multline}
where $\pt$ and $\qt$ are the transverse momenta of the quarks, $y_p$ and $y_q$ their rapidities, $\int_{\k_\perp} \equiv \int \ud^2 \k_\perp / (2\pi)^2$ and $\da\equiv \nc^2-1$ is the dimension of the adjoint representation of SU($\nc$). The expression for the ``hard matrix element'' $\Xi_{\rm coll}$ can be found in Ref.~\cite{Fujii:2013gxa}. The longitudinal momentum fractions probed in the projectile and the target, $x_1$ and $x_2$, are
\begin{equation}
	x_{1,2}=\frac{\sqrt{P_\perp^2+M^2}}{\sqrt{s}}e^{\pm Y} \; .
\end{equation}
The propagation of the $c\bar{c}$ pair in the color field of the target is described by the function
\begin{equation}\label{eq:defphi}
\phi_{_Y}^{q \bar{q},g}(\lt,\kt)=
\int\der^2 \bt \frac{N_c \, \l^2_\perp}{4 \alpha_s} \; 
S_{_Y}(\kt) \;
S_{_Y}(\lt-\kt) \;,  
\end{equation}
where $\bt$ is the impact parameter.
The function $S_{_Y}(\kt)$ is the fundamental representation dipole correlator 
in the color field of the target and it contains all the information about the target. It reads
\begin{equation}
S_{_Y}(\kt) = \int \ud^2 \rt  
e^{i\kt \cdot \rt }
S_{_Y}(\rt) \;,
\end{equation}
with
\begin{equation}
S_{_Y}(\xt-\yt) = \frac{1}{\nc }\left< \tr U^\dag(\xt)U(\yt)\right>,
\end{equation}
where $U(\xt)$ is a fundamental representation Wilson line in the color field of the target.

In the case of a proton target, where there is no explicit dependence of the dimensionless dipole amplitude on the impact parameter, the following replacement is made:
\begin{equation}\label{eq:defsigma0}
\int\der^2 \bt \to \frac{\sigma_0}{2} \; ,
\end{equation}
where $\sigma_0/2$ corresponds to the transverse area of the proton measured in DIS experiments. The function $\phi_{_p,_Y}^{q \bar{q},g}$ then reads in this case
\begin{equation}
\phi_{_p,_Y}^{q \bar{q},g}(\lt,\kt)
=
\frac{\sigma_0}{2} \; \frac{N_c \, \l^2_\perp}{4 \alpha_s} \; 
S_{_Y}(\kt) \;
S_{_Y}(\lt-\kt).
\end{equation}

For the description of the gluon distribution in the projectile proton $G_p(x_1,Q^2)$, treated in the collinear approximation, we use the MSTW 2008~\cite{Martin:2009iq} LO parametrization since the remaining of our calculation is done at leading order.

\section{Dipole correlator}
\label{sec:dipole}

For the initial condition to the running coupling Balitsky-Kovchegov equation~\cite{Balitsky:1995ub,Kovchegov:1999ua,Balitsky:2006wa} governing the rapidity evolution of $S_{_Y}(\rt)$ we use the MV$^e$ parametrization from Ref.~\cite{Lappi:2013zma} at initial rapidity $Y=\ln(1/x_0)$ with $x_0=0.01$:
\begin{equation}\label{eq:icp}
S_{Y= \ln \frac{1}{x_0}}(\rt) = \exp \left[ -\frac{\rt^2 \qso^2}{4} \ln \left(\frac{1}{|\rt| \lqcd}\!+\!e_c \cdot e\right)\right] ,
\end{equation}
and the following expression for the running coupling:
\begin{equation}
\as(r) = \frac{12\pi}{(33 - 2N_f) \log \left(\frac{4C^2}{r^2\lqcd^2} \right)} \; .
\end{equation}
The parameters $\qso^2= 0.060$ GeV$^2$, $C^2= 7.2$, $e_c=18.9$  and $\sigma_0/2 = 16.36$ mb are obtained by fitting HERA DIS data~\cite{Aaron:2009aa} for $Q^2<50$ GeV$^2$ and $x<0.01$. One advantage of this parametrization over the AAMQS~\cite{Albacete:2010sy} one, similar to the one used in Ref.~\cite{Fujii:2013gxa}, is that $S_{_Y}(\kt)$ is positive definite at any rapidity, which is necessary to interpret it as an unintegrated gluon distribution. Nevertheless we note that at LHC energies for which we will show results here, the dependence on the precise form of the initial condition is rather weak.

While the dipole correlator for a proton target can be obtained by fits to DIS data, the extension to a nucleus needs to be based on some assumptions. In Ref.~\cite{Fujii:2013gxa} the dipole correlator for nucleus was obtained by using the same initial condition as in the proton case, but with an initial saturation scale scaled by a factor $\sim A^{1/3}$. Here we instead use the Glauber approach, in which the initial condition is given by
\begin{multline}\label{eq:ica}
S^A_{Y=\ln \frac{1}{x_0}}(\rt,\bt) = \exp\bigg[ -A T_A(\bt) 
\frac{\sigma_0}{2} \frac{\rt^2 \qso^2}{4} 
\\ \times
\ln \left(\frac{1}{|\rt|\lqcd}+e_c \cdot e\right) \bigg] \; ,
\end{multline}
where the transverse thickness function $T_A$ is given by the standard Woods-Saxon distribution
\begin{equation}
T_A(\bt)= \int dz \frac{n}{1+\exp \left[ \frac{\sqrt{\bt^2 + z^2}-R_A}{d} \right]} \; ,
\end{equation}
with $d=0.54\,\mathrm{fm}$ and $R_A=(1.12A^{1/3}-0.86A^{-1/3})\,\mathrm{fm}$. Here $n$ is a normalization factor fixed so that the integral of $T_A$ over the impact parameter is unity.
To compute observables with nuclear targets we use the expressions given in the previous section but instead of making the replacement~(\ref{eq:defsigma0}) we integrate explicitly over the impact parameter. An issue in this approach is that the nucleus becomes dilute close to its edge, with a saturation scale being smaller than the one of the proton. In this region, which has a small contribution to the total cross section, we use the proton-proton result scaled so that $R_\text{pA}=1$. In this treatment the only additional input when going from a proton to nucleus target is thus the standard nuclear density $T_A$.

\section{Results}
\label{sec:results}

\begin{figure}[tb]
	\centering\includegraphics[width=0.48\textwidth]{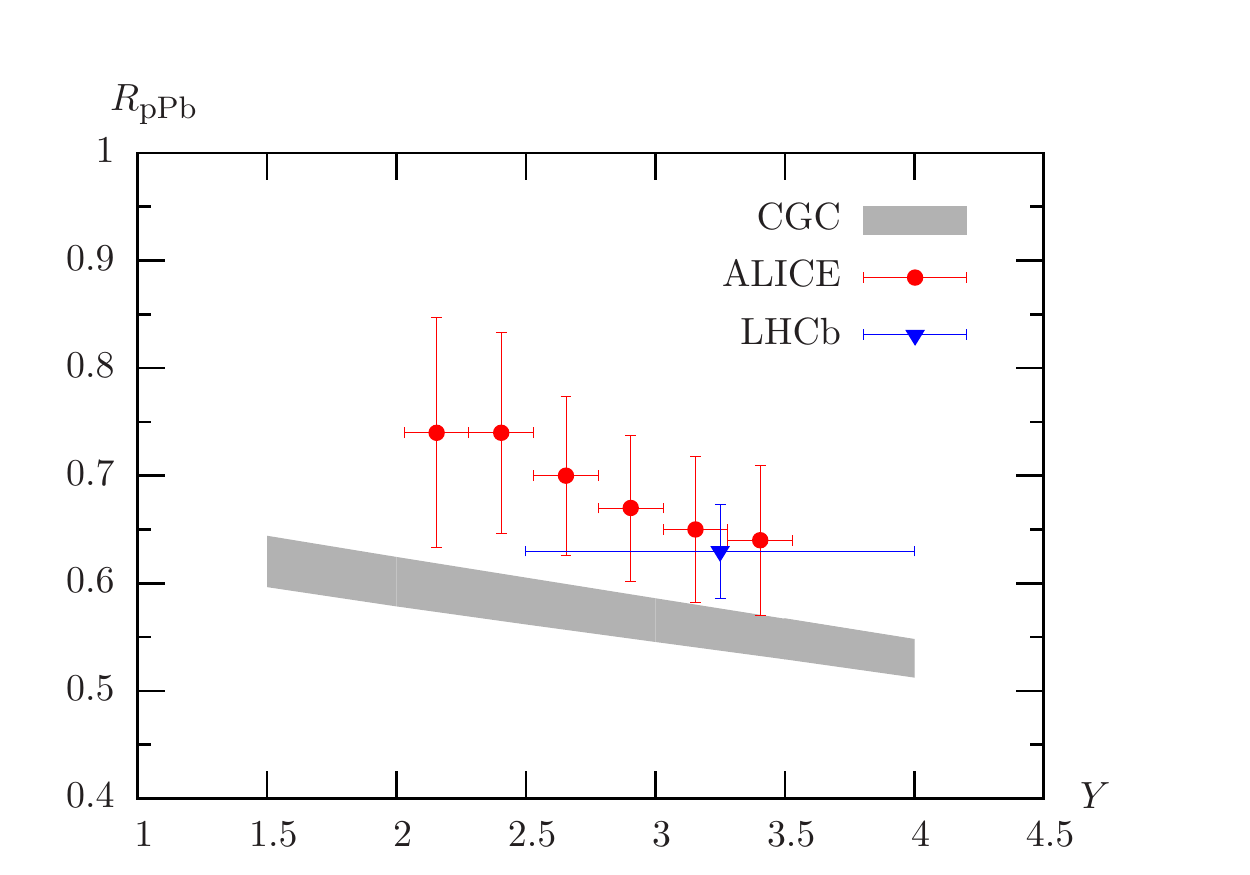}
	\caption{Nuclear modification factor for $\Jpsi$ production as a function of $Y$. Data from Refs.~\cite{Abelev:2013yxa,Aaij:2013zxa}.}
	\label{fig:RpA_Y_Jpsi}
\end{figure}

\begin{figure}[tb]
	\centering\includegraphics[width=0.48\textwidth]{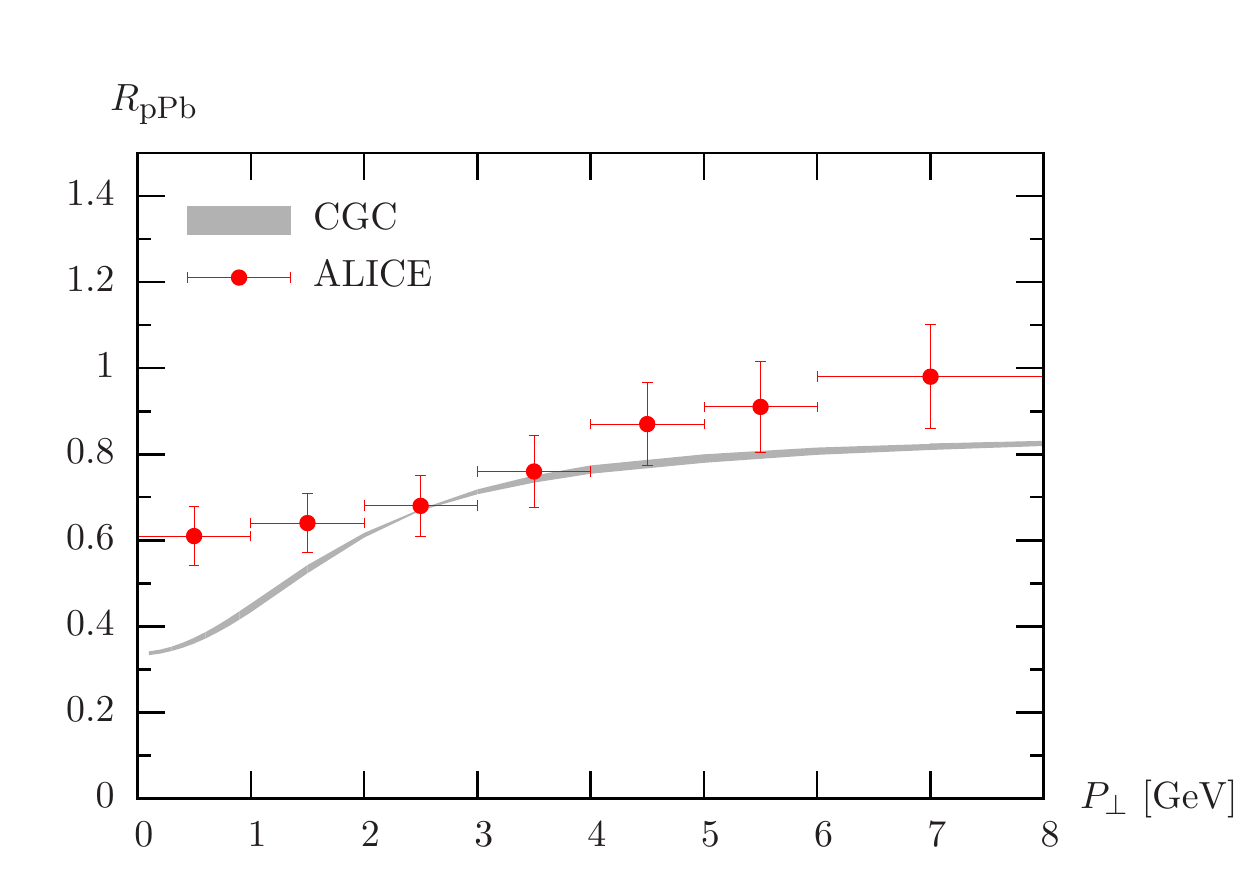}
	\caption{Nuclear modification factor for $\Jpsi$ production as a function of $P_\perp$ ($2<Y<3.5$). Data from Ref.~\cite{Abelev:2013yxa}.}
	\label{fig:RpA_pT_Jpsi}
\end{figure}

We now turn to our results. We will here focus on the nuclear modification factor $R_\text{pA}$; for results on the cross sections for proton-proton and proton-nucleus collisions we refer to Ref.~\cite{Ducloue:2015gfa}, where it was observed that the general shapes for the cross sections as a function of $P_{\perp}$ and $Y$ are in quite good agreement with experimental data, although they are affected by a quite large normalization uncertainty. These uncertainties cancel to a large extend in $R_\text{pA}$, as can be observed from Figs.~\ref{fig:RpA_Y_Jpsi} and~\ref{fig:RpA_pT_Jpsi} where we show $R_\text{pPb}$ at $\sqrt{s_{NN}}=5$~TeV as a function of $Y$ and $P_{\perp}$ respectively. The uncertainty band in both figures corresponds to the variation of the charm quark mass between 1.2 and 1.5 GeV and of the factorization scale between $M_\perp/2$ and $2M_\perp$ with $M_\perp=\sqrt{M^2+P_{\perp}^2}$ and $M$ is the $c\bar{c}$ pair's invariant mass. The comparison with measurements at the LHC by the ALICE and LHCb collaborations shows that the values obtained here are much closer to the data than the previous work in Ref.~\cite{Fujii:2013gxa}, but they are still slightly too small to describe the data, especially at low $P_{\perp}$ as can be seen from Fig.~\ref{fig:RpA_pT_Jpsi}.

\section{Conclusions}
\label{sec:disc}

In this work we have re-evaluated the CGC predictions for the production of $\Jpsi$ mesons in high energy proton-proton and proton-nucleus collisions, in particular focusing on the treatment of the nuclear geometry when going from a proton target to a nuclear one. The dipole cross section in the proton case is fully constrained by DIS fits, while the only additional input used for a nucleus target in this approach is the standard Woods-Saxon distribution. Our results in this Glauber approach for the nuclear modification factor, which has a quite small uncertainty, are closer to experimental data than previous estimates. They nevertheless slightly underestimate this ratio, especially at low transverse momentum.

\section*{Acknowledgements}
B.~D. and T.~L. are supported by the Academy of Finland, projects 
267321 and 273464 and H.M by 
the Graduate School of Particle and Nuclear Physics.
This work was done using computing resources from
CSC -- IT Center for Science in Espoo, Finland.

\end{document}